\documentclass[aps,prd,twocolumn,preprintnumbers,showpacs,superscriptaddress,nofootinbib]{revtex4}
\usepackage{graphicx,amssymb,amsmath}
\def\fsl#1{\setbox0=\hbox{$#1$}                 
   \dimen0=\wd0                                 
   \setbox1=\hbox{/} \dimen1=\wd1               
   \ifdim\dimen0>\dimen1                        
      \rlap{\hbox to \dimen0{\hfil/\hfil}}      
      #1                                        
   \else                                        
      \rlap{\hbox to \dimen1{\hfil$#1$\hfil}}   
      /                                         
   \fi}                                         %

\newcommand{\tr}{\mbox{tr}}
\newcommand{\VEV}[1]{\langle #1 \rangle}

\begin{document}

\date{\today}

\title{Topcolor Dynamics and \\ 
The Effective Gluon-Gluon-Higgs Operator}

\author{Michio Hashimoto}

\email[E-mail:]{michioh@post.kek.jp}

\affiliation{Theory Group, KEK,
Oho 1-1, Tsukuba, Ibaraki 305-0801, Japan}


\begin{abstract}
We discuss the production of the composite Higgs boson 
in topcolor models via the gluon fusion process.
We consider the contribution of color-octet massive gauge bosons 
(colorons) strongly interacting with the top quark, 
in addition to nonstandard contributions of 
the top-Yukawa coupling and heavy colored fermions 
other than the top quark. 
In order to estimate the contribution of colorons, 
we derive the low-energy effective theory by eliminating colorons 
by using the equation of motion for colorons. 
We replace the composite operator $(\bar{q}_L t_R)(\bar{t}_R q_L)$
in the effective theory by the composite Higgs operator. 
We then obtain the effective gluon-gluon-Higgs ($ggH$-) operator
induced by colorons and find that its coefficient 
(${\cal A}_{\rm col}$) is proportional to $m_{\rm dyn}^2/M^2$, 
where $M$ and $m_{\rm dyn}$ denote the coloron mass and 
the mass dynamically generated by colorons, respectively.
The contribution of colorons ${\cal A}_{\rm col}$ becomes 
comparable to the top-loop effect ${\cal A}_{\rm top}$ 
for $M \sim {\cal O}(\rm 1 TeV)$ and 
$m_{\rm dyn} \sim {\cal O}(0.6 \rm TeV)$.
Such a large dynamical mass can be realized in top-seesaw (TSS) models 
consistently with the experimental value of the top quark mass 
($m_t^{(\rm exp)}$),
while the dynamical mass itself is adjusted to $m_t^{(\rm exp)}$ 
in topcolor assisted technicolor models (TC2).
We find that the coloron contribution ${\cal A}_{\rm col}$ 
can be sizable in a certain class of TSS models: 
the contribution of colorons (the top-loop) is dominant 
in the real (imaginary) part of the $H \to gg$ amplitude
for the Higgs boson mass $m_H$ of the order of 1 TeV.
On the other hand, enhancement of the top-Yukawa coupling becomes
important in TC2.
We can observe signatures of the Higgs boson in TC2 
with $m_H \sim 200$ GeV 
even at the Tevatron Run II 
as well as at the LHC.
We estimate $S/\sqrt{B}=3-6$ for an integrated luminosity
of 2 fb${}^{-1}$ and $m_H=190$ GeV at the Tevatron Run II. 
\end{abstract}

\maketitle

\section{Introduction}

The gauge interaction properties of the Standard Model (SM) 
have been measured quite precisely in the last decade. 
However, the Higgs particle has not yet been discovered in spite of 
much effort.
The physics behind the electroweak symmetry breaking 
(EWSB) and the origin of masses of quarks and leptons are left as
unresolved problems. 
The idea of the top quark condensate~\cite{MTY89, Nambu89},
in which the 4-top-quark interaction is introduced 
to trigger the EWSB,
explains naturally the large mass of the top quark at 
the EWSB scale. 
(See also the earlier attempt~\cite{Terazawa}.)
This model is often called the ``top mode standard model'' (TMSM),
because the scalar bound state of $\bar{t}t$ plays the role of 
the Higgs boson in the SM.

It is known that the original version of the TMSM has 
some difficulties: 
the top quark mass $m_t$ is predicted
about 10\%--30\% larger than the experimental value 
($m_t^{({\rm exp})} = 174$ GeV),
even if we take the ultraviolet cutoff 
(or the compositeness scale) 
to the Planck or the GUT scale~\cite{MTY89,BHL,MH98}. 
In addition, such a huge cutoff causes a serious fine-tuning 
problem.
In topcolor models (TCMs)~\cite{topcolor,TC2,topseesaw}, 
the 4-top-quark interaction, whose origin is not specified in 
the original version of the TMSM, is provided by 
exchange of colorons which are color-octet massive gauge bosons
strongly interacting with the top quark. \cite{review}
If we assume that the coloron mass $M$ is 
${\cal O}(1 \rm TeV)$, 
we need not tune the coupling of the 4-top-quark finely to 
its critical value.
Although the mass generated by the dynamics of 
colorons (the dynamical mass $m_{\rm dyn}$) becomes quite large, 
typically $m_{\rm dyn} = 0.6-0.7$ TeV, 
in TCMs where only the top-condensate is responsible for the EWSB,
the problem $m_{\rm dyn} \gg m_t^{({\rm exp})}$ 
can be resolved in some classes of models such as 
topcolor assisted technicolor models (TC2)~\cite{TC2} 
and top-seesaw (TSS) models~\cite{topseesaw}.
In TC2, we introduce techni-fermions in addition to colorons.
We assume that 
the topcolor interaction is responsible for the top-quark 
mass, while the EWSB occurs mainly due to technicolor interactions.
Namely, the dynamical mass itself is adjusted to 
$m_t^{({\rm exp})}$ with 
a small vacuum expectation value (VEV) of the top quark condensate
$v_{\rm top} \simeq v/(3-4)$, 
where $v$ denotes the EWSB scale.
The top-Yukawa coupling in TC2 is about $3-4$ times larger than 
the value in the SM.
This enhancement of the top-Yukawa is quite important to 
the Higgs boson production via the gluon fusion process. 
In TSS models, we introduce 
a vector-like heavy colored fermion $\chi$, 
whose mass term has nothing to do with the EWSB,
in order to have a seesaw-type mass matrix.
The dynamical mass $m_{\rm dyn}$ in TSS models does not correspond to 
the mass eigenvalue of the top-quark.
The experimental value $m_t^{({\rm exp})}$ is obtained 
after diagonalization of the mass matrix for the top-quark
via the seesaw mechanism.
Namely, the large dynamical mass of the order of 0.6 TeV
can be realized in TSS models consistently with $m_t^{({\rm exp})}$. 
Here, we note that the mixing between $t_L$ and $\chi_L$ is 
severely constrained from the custodial symmetry violation, i.e.
the $T$-parameter, which usually requires 
$\chi$ to have a mass of the order of several-TeV.
Although the TMSM with extra dimensions proposed by 
Arkani-Hamed et al.~\cite{ACDH} is also a candidate to resolve 
the fine-tuning problem etc., 
it is a pure bulk gauge theory without 4-fermion interactions 
in the bulk rather than a TCM~\cite{TMSM_ED}.
The phenomenology of the TMSM with extra dimensions will be studied 
elsewhere~\cite{future}.

In this paper, we study the production of the Higgs boson 
in TCMs. 
We regard the Higgs boson as the tightly bound state of $\bar{t}t$
(or $\bar{t}\chi$) 
and do not consider direct signatures of other bound states. 
The Higgs boson can be searched for
at the Tevatron~\cite{tevatron} or at the LHC~\cite{LHC}.
In the SM, the leading contribution to the Higgs boson production 
at hadron colliders comes from the gluon fusion process 
via loops of heavy quarks~\cite{Georgi:1978gs}.
The gluon fusion process is thus sensitive to the top-Yukawa coupling 
and the number of heavy quarks.
In addition, it is possible that extensions to QCD affect 
the gluon fusion process.
Especially, colorons are strongly coupled to the top quark, 
and the dynamical mass $m_{\rm dyn}$ is very large for TCMs
with a single VEV triggering the EWSB, 
so that the effect of colorons can be expected to be sizable. 
In order to evaluate roughly the size of the contribution of colorons,
we first derive the low-energy effective theory 
by eliminating colorons by using the equation of motion (EOM) 
for colorons.
We then find that the effective gluon-gluon-Higgs- ($ggH$-) operator 
is induced in the low-energy effective theory and that 
the coefficient is proportional to $m_{\rm dyn}^2/M^2$. 
In the situation of $M \sim {\cal O}({\rm 1 TeV})$ and 
$m_{\rm dyn} \sim {\cal O}(0.6 \rm TeV)$, 
the contribution of colorons becomes comparable to 
the top-loop effect.
Next, we estimate quantitatively the size of 
the effective $ggH$-coupling including 
the contributions of the top-quark loop, 
other heavy-quark loops, and colorons in TSS models and TC2.
We find that the contribution of colorons can be
sizable in TSS models with the coloron mass 
$M \sim {\cal O}(\rm 1 TeV)$
and the mass of $\chi$, $m_\chi \sim {\cal O}(10 \rm TeV)$,
and that the effect of the top-loop is almost same as in the SM.
We estimate numerically the effects of colorons and the top-loop as 
${\cal A}_{\rm col}=[-(1-3)+(0.1-0.2)i] \times 10^{-2}
 (\rm TeV^{-1})$ and
${\cal A}_{\rm top}=[(0.4-0.1)+(1.4-1.1)i] \times 10^{-2}
 (\rm TeV^{-1})$ for $m_H=0.8-1$ TeV, respectively. 
Namely, the contribution of colorons (the top-loop) is dominant
in the real (imaginary) part of the $H \to gg$ amplitude.
In TC2, the contribution of colorons becomes small, while
the effect of the top-loop is strongly enhanced.
We find that signatures of the Higgs boson in TC2 
can be observed even at the Tevatron Run II as well as at the LHC.
In particular, we evaluate $S/\sqrt{B}=3-6$ for 
an integrated luminosity of 2 fb$^{-1}$ 
and $m_H=190$ GeV at the Tevatron Run II.

This paper is organized as follows:
In Sec.~II we derive the low-energy effective theory by eliminating 
colorons by using the EOM in a TCM with only the top- and 
bottom-quarks, for illustration. 
We find that the contribution of colorons is proportional to 
$m_{\rm dyn}^2/M^2$. 
In Sec.~III, we estimate quantitatively the size of the effective
$ggH$-coupling in TSS models and in TC2.
We find that the contribution of colorons can be sizable 
in TSS models, while it cannot in TC2.
As for the effect of the top-loop, it is very large in TC2. 
We show expected signals of the Higgs boson in TC2 
at the Tevatron Run II and at the LHC.
Sec.~IV is devoted to summary and discussion.

\section{The effective $\mbox{\boldmath $ggH$}$-operator 
induced by colorons}

We define the amplitude of $H \to gg$ as
\begin{widetext}
\begin{equation}
 A(H \to G_\mu^a(p)G_\nu^{a'}(k)) \equiv -4\delta^{aa'}
 \epsilon_\mu^*(p)\epsilon_\nu^*(k)((p \cdot k)g^{\mu\nu}-k^\mu p^\nu) 
 {\cal A}, \label{ggH_amp}
\end{equation}
\end{widetext}
where $G_\mu^a (p)$ denotes a gluon with momentum $p$
and the suffix $a$ is the index of colors 
for the adjoint representation. 
In the SM, we can obtain ${\cal A}={\cal A}_{\rm SM}$
at the 1-loop level,
\begin{equation}
  {\cal A}_{\rm SM}(\tau) = 
  \frac{\alpha_s}{8\pi v} \tau (1-(\tau-1)f(\tau)) \label{sm_finite}
\end{equation}
with $v=\mbox{246 GeV}$, where we have defined 
$\tau \equiv 4m_t^2/m_H^2$ and
\begin{equation}
  f(\tau) \equiv \left\{
  \begin{array}{lc}
  {\rm arcsin}^2\frac{1}{\sqrt{\tau}}, & (\tau \geq 1) \\
  -\frac{1}{4}\left(\ln \frac{1+\sqrt{1-\tau}}{1-\sqrt{1-\tau}}
                    -i\pi\right)^2,  & (\tau < 1)
  \end{array}\right. . \label{f}
\end{equation}
Since the factor ${\cal A}_{\rm SM}$ in the $\tau \to \infty$ limit 
becomes independent on $\tau$,
\begin{equation}
  {\cal A}_{\rm SM}(\infty)=\frac{\alpha_s}{12\pi v}, \label{sm_inf}
\end{equation}
the $H \to gg$ amplitude can be written as the local operator 
\begin{equation}
  {\cal O}_{ggH} = 
  {\cal A}_{\rm SM}(\infty) H G_{\mu\nu}^a G^{a \mu\nu}.
  \label{sm}  
\end{equation}
In this sense,
the factor ${\cal A}$ 
corresponds to the coefficient of the local operator 
${\cal O}_{ggH}$.
We also note that the $H \to gg$ amplitude has the imaginary part
above the threshold of the top-pair production ($m_H > 2 m_t$).

The gluon fusion process is obviously sensitive to
the top-Yukawa coupling and the number of heavy quarks. 
In this paper, we also consider the effect of colorons 
strongly interacting with the top quark.
We hence classify nonstandard contributions 
to the $H \to gg$ amplitude into three categories:
\begin{description}
\item[(a)] {\it Loop effects of the top quark;}\\
In the SM, the lowest order amplitude for the gluon fusion 
arises from the triangle diagram of the top quark.
On the other hand,
the top-Yukawa coupling can be larger than the SM one
in TCMs with multi composite Higgs doublets.
The enhancement of the top-Yukawa coupling is quite important 
for the gluon fusion process.

\item[(b)] {\it Loop effects of other heavy quarks;}\\
Generally, heavy quarks other than the top are also introduced 
in TCMs. 
They contribute to the $H \to gg$ amplitude at the 1-loop level. 

\item[(c)] {\it Contributions of strongly interacting colorons;}\\
Since colorons are strongly coupled to the top quark,
effects of colorons are potentially sizable.
\end{description}
Corresponding to the three contributions (a), (b) and (c),
we split the factor ${\cal A}$ in Eq.~(\ref{ggH_amp})
into three parts,
${\cal A}_{\rm top}$, ${\cal A}_{\rm heavy}$ and ${\cal A}_{\rm col}$,
\begin{equation}
  {\cal A} = 
  {\cal A}_{\rm top} + {\cal A}_{\rm heavy} + {\cal A}_{\rm col}.
 \label{all_A}
\end{equation}

In this section, we roughly estimate the size of (c).
For this purpose, we derive the low-energy effective theory 
by eliminating colorons by using the equation of motion (EOM) 
for colorons.
For illustration, we take a toy model of TCMs with only the top- and
bottom-quarks. 
This model is not realistic, but it is sufficient to understand
the origin of the coloron contribution ${\cal A}_{\rm col}$. 
In the next section, we evaluate quantitatively the sizes of 
${\cal A}_{\rm top}, {\cal A}_{\rm heavy}$, and ${\cal A}_{\rm col}$
in typical TCMs such as TC2 and TSS models.

After the topcolor symmetry is spontaneously broken down, 
we write down the model,
\begin{widetext}
\begin{equation}
{\cal L} = {\cal L}_{\rm top}+{\cal L}_{\rm int}+{\cal L}_{\rm col},
\label{TCM}
\end{equation}
with
\begin{eqnarray}
{\cal L}_{\rm top} &\equiv& 
   \bar{q}_L i\fsl{D} q_L + \bar{t}_R i\fsl{D}t_R + 
   \bar{b}_R i\fsl{D}b_R 
    -\frac{1}{2}\tr(G_{\mu\nu}G^{\mu\nu}) , \label{L_top} \\[1mm]
  {\cal L}_{\rm int} &\equiv&  g' A^a_\mu (J_L^{a\mu}+J_R^{a\mu}) , 
  \label{L_int} \\
  {\cal L}_{\rm col} &\equiv& 
  - \frac{1}{2}\tr( A_{\mu\nu} A^{\mu\nu})
  + M^2 \tr(A_\mu A^\mu)  \nonumber \\
 && + i g_{GAA} \tr(G_{\mu\nu}[A^\mu,A^\nu]) 
    + i g_{A^3} \tr(A_{\mu\nu}[A^\mu,A^\nu]) 
    + g_{A^4} \tr([A^\mu,A^\nu]^2) \label{L_col}
\end{eqnarray}
up to dimension four operators, 
where $g'$ is the coupling constant between the top quark and 
colorons, and 
$A_\mu$ and $G_\mu$ denote the coloron field and the gluon, 
respectively~\footnote{
In order to avoid the bottom quark condensation, 
we have to introduce a new strong $U(1)$-gauge interaction, 
for example.
Instead of specifying the manner for suppression of 
the bottom condensation, 
we adjust the 4-fermion interactions so that 
only the top condensation takes place.}.
The current $J_{L(R)}^{a \mu}$ for the weak doublet $q_L$ 
(the weak singlet $t_R$, $b_R$) is defined as usual. 
We note that $G_\mu$ and $A_\mu$ are matrices defined as
$G_\mu \equiv G_\mu^a T^a, A_\mu \equiv A_\mu^a T^a$ with 
generators $T^a$ of $SU(N_c)$, where $N_c$ is the number of colors. 
In Eqs.~(\ref{L_top})--(\ref{L_col}) with the gauge symmetry 
of QCD, 
the covariant derivative $D_\mu$, and the field strength 
$G_{\mu\nu}$ and $A_{\mu\nu}$ for the gluon and the coloron 
are written as
\begin{equation}
 D_\mu \equiv \partial_{\mu}-ig_s G_\mu , \quad 
 G_{\mu\nu} \equiv \frac{i}{g_s}[D_{\mu},D_{\nu}], \quad
 A_{\mu\nu} \equiv 
 \partial_\mu A_\nu - \partial_\nu A_\mu 
 -ig_s[G_\mu,A_\nu]+ig_s[G_\nu,A_\mu] .
\end{equation}
We do not incorporate higher dimensional operators in 
${\cal L}_{\rm col}$. 
If we would allow them, we could freely choose the coefficient 
${\cal A}_{\rm col}$ of the operator ${\cal O}_{ggH}$, 
as we will see later on.
We note that the triple coupling $g_{GAA}$ plays an important role
in the following analysis and that it is equal to the QCD coupling,
\begin{equation}
 g_{GAA} = g_s, 
\end{equation}
in usual TCMs. 

Now, we derive the effective theory at the coloron scale
in the classical approximation.
We eliminate colorons from Eq.~(\ref{TCM})
by using the EOM for $A_\mu$,
\begin{eqnarray}
  M^2 A_{\mu} &=& 
  -g' J_\mu +[D^\nu,A_{\mu\nu}]-ig_{GAA}[A^\nu,G_{\mu\nu}] 
  \nonumber \\ &&
   -ig_{A^3}[D^\nu,[A_\mu,A_\nu]]-ig_{A^3}[A^\nu,A_{\mu\nu}]
  -g_{A^4}[A^\nu,[A_\mu,A_\nu]]  \label{EOM}
\end{eqnarray}
with $J^\mu \equiv J_L^\mu + J_R^\mu$,
where we can expand $A_\mu$ in powers of $1/M^2$
with recursive use of Eq.~(\ref{EOM}) and can rewrite $A_\mu$
in terms of $D_\mu$, $J_\mu$ and $G_{\mu\nu}$:
\begin{eqnarray}
\lefteqn{
A_{\mu}=-\frac{g'}{M^2} J_\mu } \nonumber \\ &&
  +\frac{g'}{(M^2)^2}ig_{GAA}[J^\nu,G_{\mu\nu}] 
  -\frac{g'}{(M^2)^2}[D_\nu,[D_\mu,J^\nu]]
  +\frac{g'}{(M^2)^2}[D_\nu,[D^\nu,J^\mu]] \nonumber \\ &&
  +\frac{g'}{(M^2)^3}g_{GAA}^2
   [[J_\lambda,G^{\nu\lambda}],G_{\mu\nu}] 
  +\frac{g'}{(M^2)^3}ig_{GAA}
   [[D^\lambda,[D^\nu,J_\lambda]],G_{\mu\nu}] 
  -\frac{g'}{(M^2)^3}[D_\nu,[D_\mu,[D_\lambda,[D^\nu,J^\lambda]]]]
  \nonumber \\ &&
  +\frac{g'}{(M^2)^3}ig_{GAA}
   [D_\nu,[D_\mu,[J_\lambda,G^{\nu\lambda}]]] 
  +\frac{g'}{(M^2)^3}[D_\nu,[D_\mu,[D_\lambda,[D^\lambda,J^\nu]]]]
  +\frac{g'}{(M^2)^3}[D_\nu,[D^\nu,[D_\lambda,[D_\mu,J^\lambda]]]]
  \nonumber \\ &&
  -\frac{g'}{(M^2)^3}[D_\nu,[D^\nu,[D_\lambda,[D^\lambda,J_\mu]]]]
  -\frac{g'}{(M^2)^3}ig_{GAA}[D_\nu,[D^\nu,[J^\lambda,G_{\mu\lambda}]]]
  -\frac{g'}{(M^2)^3}ig_{GAA}
   [[D_\lambda,[D^\lambda,J^\nu]],G_{\mu\nu}] \nonumber \\ &&
  -\frac{(g')^2}{(M^2)^3}ig_{A^3}[D_\nu,[J_\mu,J^\nu]]
  -\frac{(g')^2}{(M^2)^3}ig_{A^3}[J_\nu,[D_\mu,J^\nu]] 
  +\frac{(g')^2}{(M^2)^3}ig_{A^3}[J_\nu,[D^\nu,J_\mu]] \nonumber \\ &&
  + {\cal O}\left(\frac{1}{(M^2)^4}\right)
\end{eqnarray}
Noting that $J_\mu$ is the conserved current, i.e.
$[D_\mu,J^\mu]=\partial_\mu J^\mu -ig_s[G_\mu,J^\mu] =0$, we find 
\begin{equation}
A_{\mu}=-\frac{g'}{M^2} J_\mu
  +\frac{g'}{(M^2)^2}i(g_{GAA}+g_s)[J^\nu,G_{\mu\nu}]
  +\frac{g'}{(M^2)^3}(g_{GAA}+g_s)^2
   [G_{\mu\nu},[G^{\nu\lambda},J_\lambda]] + \cdots, \label{EOM2}
\end{equation}
where we neglected irrelevant terms to the effective $ggH$-coupling.
Substituting Eq.~(\ref{EOM2}) for 
Eqs.~(\ref{L_int}) and (\ref{L_col}), 
we obtain the effective Lagrangian 
written in terms of local composite operators of $J_\mu$
and $G_\mu$: 
\begin{equation}
{\cal L}_{\rm int}+{\cal L}_{\rm col} \to {\cal L}_{\rm comp}
\end{equation}
with
\begin{equation}
{\cal L}_{\rm comp} \equiv
 -\frac{g'{}^2}{2M^2} J^{a\mu}J_\mu^a 
 -\frac{(g_{GAA}+g_s) g'{}^2}{2(M^2)^2}
             f_{abc}G_{\mu\nu}^a J^{b\mu}J^{c\nu} 
   -\frac{(g_{GAA}+g_s)^2 g'{}^2}{2(M^2)^3}
          f_{abe}f_{cde}G_{\mu\nu}^a J^{b\nu}
                        G^{c\,\mu\lambda} J_\lambda^d + \cdots .
 \label{eff_Lag}
\end{equation}
Since only scalar operators with large anomalous dimension
are relevant in the low-energy effective theory,
we neglect vectorial and tensorial terms of 4-fermion operators.
The part including the 4-top-quark operator is given by
\begin{equation}
  {\cal L}_{\rm 4-top} = 
     \frac{g_{t0}^2}{M^2} 
    (\bar q_L t_R)(\bar t_R q_L)
    + \frac{N_c^{-1}}{4}\frac{(g_{GAA}+g_s)^2 g'{}^2}{(M^2)^3}
     G_{\mu\nu}^a G^{a\,\mu\nu} (\bar q_L t_R)(\bar t_R q_L) ,
    \quad (g_{t0} \gtrsim g')
    \label{ggH0} 
\end{equation}
at the leading order of the $1/N_c$-expansion, 
where we have used the Fierz transformation.
The first term in Eq.~(\ref{ggH0}) is the driving force of 
the top condensation, while
the second one is the source of the $ggH$-operator.
(See also Fig.~\ref{gg4Fermi}.)
Here, we note that
the coefficient of the 4-top-quark (4-bottom-quark) operator 
is generally required to be larger (smaller) than $(g')^2$, which 
is nearly equal to the critical coupling $(g_{\rm crit})^2$,
in order to produce the desirable pattern of condensations. 
We thus replaced the coefficient by $g_{t0}^2$ ($g_{t0} \gtrsim g'$)
in Eq.~(\ref{ggH0}), assuming a certain mechanism for suppression 
of the bottom condensation. 
%
%
\begin{figure}[tbp]
  \begin{center}
    \resizebox{0.8\textwidth}{!}{\includegraphics{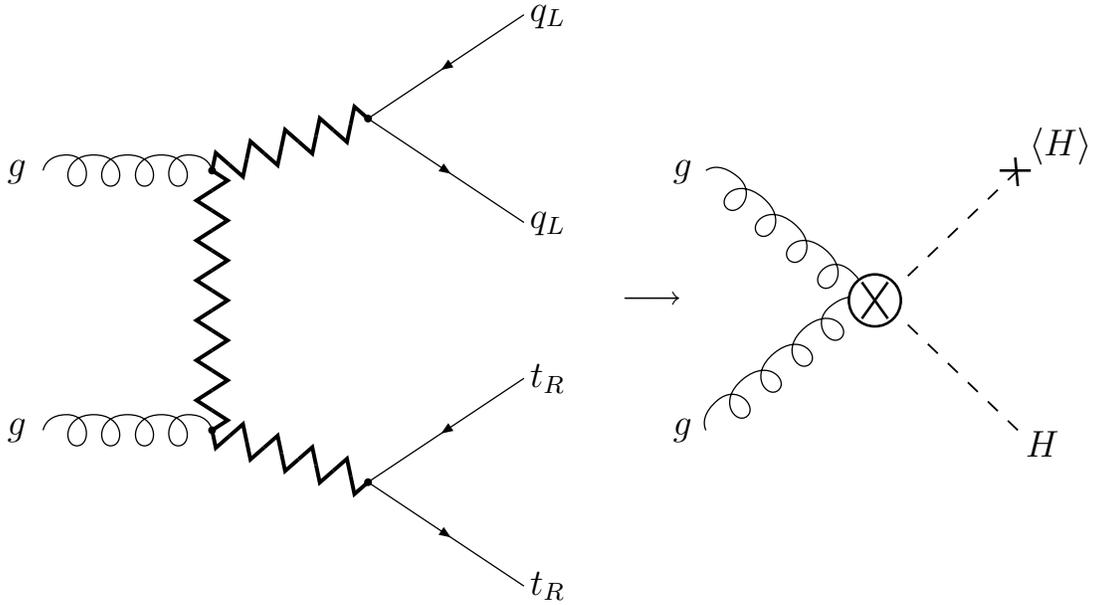}}
  \end{center}
 \caption{The effective gluon-gluon-Higgs operator induced by colorons.
          Internal zigzag lines represent colorons. 
          After the Fierz transformation, we replace the 
          4-top-quark operator $(\bar{q}_L t_R)(\bar{t}_R q_L)$ by 
          the composite Higgs field. \label{gg4Fermi}}
\end{figure}

In order to obtain the low-energy effective theory with 
the Higgs doublet $\phi$,
we rewrite the effective Lagrangian Eq.~(\ref{ggH0}) 
at the coloron scale as follows:
\begin{eqnarray}
  {\cal L}_{\phi_0} & = &
      - M^2 \phi_0^\dagger \phi_0 
      - \left[g_{t0} (\bar q_L t_R)\phi_0 + \mbox{h.c.}\right]
      \nonumber \\
  &&  -\frac{N_c^{-1}}{4}\frac{g'{}^2}{g_{t0}^2}
       \frac{(g_{GAA}+g_s)^2}{M^2}
       G_{\mu\nu}^a G^{a\,\mu\nu} \phi_0^\dagger \phi_0 
      -\frac{N_c^{-1}}{4}\frac{g'{}^2}{g_{t0}^2}
       \frac{(g_{GAA}+g_s)^2}{(M^2)^2}
       G_{\mu\nu}^a G^{a\,\mu\nu}
       \left( g_{t0}(\bar q_L t_R)\phi_0 + \mbox{h.c.} \right) 
      \label{H0}
\end{eqnarray}
with the bare Higgs doublet $\phi_0$. 
We can confirm the equivalence of Eq.~(\ref{H0}) and 
Eq.~(\ref{ggH0}) through the EOM for $\phi_0^\dagger$ and $\phi_0$.
The kinetic term of the Higgs doublet develops 
below the coloron scale. 
We then obtain the low-energy effective theory at the weak scale:
\begin{equation}
  {\cal L}_{\rm eff} = {\cal L}_{\rm top} + {\cal L}_\phi
\end{equation}
with
\begin{eqnarray}
{\cal L}_\phi &\equiv & \partial_\mu \phi^\dagger \partial^\mu \phi
 - m_\phi^2 \phi^\dagger \phi
 - \lambda_\phi (\phi^\dagger \phi)^2
 - g_t\left[(\bar q_L t_R)\phi+\mbox{h.c.}\right] \nonumber \\
&&  -\frac{N_c^{-1}}{4Z_\phi}\frac{g'{}^2}{g_{t0}^2}
     \frac{(g_{GAA}+g_s)^2}{M^2}
       G_{\mu\nu}^a G^{a\,\mu\nu} \phi^\dagger \phi
      -\frac{N_c^{-1}}{4}\frac{g'{}^2}{g_{t0}^2}
       \frac{(g_{GAA}+g_s)^2}{(M^2)^2}
       G_{\mu\nu}^a G^{a\,\mu\nu}
       \left[ g_t(\bar q_L t_R)\phi + \mbox{h.c.} \right], 
      \label{H}
\end{eqnarray}
\begin{equation}
 \phi \equiv Z_\phi^{1/2}\phi_0, \quad \mbox{and} \quad
 g_t \equiv g_{t0}/Z_\phi^{1/2}. \label{g_t}
\end{equation}
Since the neutral component of $\phi$ can be written as 
$(v+H(x))/\sqrt{2}$, the coefficient ${\cal A}_{\rm col}$ 
of the effective $ggH$-operator is given by
\begin{equation}
 {\cal A}_{\rm col} = 
 - \frac{N_c^{-1}}{4} \frac{g'{}^2}{g_{t0}^2}
   \frac{(g_{GAA}+g_s)^2}{M^2}\frac{v}{Z_\phi} .
 \label{ggH}
\end{equation}
\end{widetext}
Here, we comment on the reason why we did not allow 
higher dimensional operators in the Lagrangian for colorons. 
If the dimension six operator 
$\tr([G_{\mu\nu},A^\nu][G^{\mu\lambda},A_\lambda])$ 
was not forbidden in ${\cal L}_{\rm col}$,
it would lead to the same composite operator as 
the third term of ${\cal L}_{\rm comp}$ in Eq.~(\ref{eff_Lag}).
This means that the coefficient ${\cal A}_{\rm col}$ of the operator
${\cal O}_{ggH}$ at the weak scale is not constrained.
Namely, tolerance of higher dimensional operators in 
${\cal L}_{\rm col}$ is equivalent to adding ${\cal O}_{ggH}$ 
with a free coefficient 
by hand.

The VEV of $\bar{t} t$ has a nonzero value
only when the coefficient $(g_{t0})^2$ of the 4-top quark interaction 
is larger than its critical coupling~\cite{NJL, vaks}, 
\begin{equation}
  (g_{t0})^2 \geq (g_{\rm crit})^2, 
 \quad (g_{\rm crit})^2 = \frac{8\pi^2}{N_c}. \label{crit}
\end{equation}
The top quark then acquires the dynamical mass, 
\begin{equation}
  m_{\rm dyn} \equiv \frac{g_t}{\sqrt{2}} v 
  = \frac{g_{t0}}{\sqrt{2}Z_\phi^{1/2}} v . \label{m_t}
\end{equation}
Within the usual large $N_c$-bubble approximation, 
we find the wave function renormalization of the Higgs doublet, 
\begin{equation}
 Z_\phi(\mu^2) \simeq \frac{N_c g_{t0}^2}{16\pi^2} \ln M^2/\mu^2 
\label{Z_phi}
\end{equation}
with the renormalization point $\mu$. 
Substituting Eq.~(\ref{Z_phi}) for Eq.~(\ref{m_t}), 
we obtain the relation
\begin{equation}
  v^2 = \frac{N_c}{8 \pi^2}m_{\rm dyn}^2 \ln M^2/m_{\rm dyn}^2,
 \label{PS0}
\end{equation}
which is called the Pagels-Stokar (PS) formula~\cite{PS},
and can evaluate the dynamical mass as 
$m_{\rm dyn} \sim {\cal O}(0.6 \rm TeV)$ for 
$M \sim {\cal O}(\rm 1 TeV)$.
Although the top quark mass is predicted too large 
($m_t=m_{\rm dyn}$) in this toy model, 
we consider realistic TCMs such as TSS models and TC2 
in the next section.

Now, we estimate roughly the size of ${\cal A}_{\rm col}$ and 
compare it with the size of ${\cal A}_{\rm top}$.
Using $g_{t 0} \sim g_{\rm crit}$, Eq.~(\ref{Z_phi}) 
reads $Z_\phi \sim {\cal O}(1)$ up to a logarithmic term.
Namely, the expression of ${\cal A}_{\rm col}$ in Eq.~(\ref{ggH}) 
has no loop-suppression factor $1/(4\pi)^2$.
It is, in other words, a consequence of 
the large dynamical mass obtained from the PS formula,
$m_{\rm dyn} \sim 4\pi v$ up to a logarithmic term.
Actually, we can explicitly rewrite Eq.~(\ref{ggH}) as
\begin{equation}
{\cal A}_{\rm col} = 
- \frac{(g_{GAA}+g_s)^2}{2 N_c g_{t0}^2 v}\frac{g'{}^2}{g_{t0}^2}
  \frac{m_{\rm dyn}^2}{M^2} 
\label{ggH2}
\end{equation}
in terms of $m_{\rm dyn}$, 
where 
$g_{t0} \approx g' \sim g_{\rm crit}$ and the suppression factor
$2N_c g_{t0}^2 \sim (4\pi)^2$ is canceled by 
$m_{\rm dyn} \sim 4\pi v$.
We find roughly that the contribution of colorons is 
${\cal A}_{\rm col} \sim - g_s^2 v/M^2$ with $g_{GAA}=g_s$, 
while the top-loop effect is 
${\cal A}_{\rm top} \sim g_s^2 /[(4\pi)^2 v]$ from Eq.~(\ref{sm_inf}).
In the toy model, we can estimate the contribution of 
the top quark loop as
${\cal A}_{\rm top}={\cal A}_{\rm SM}(1)=\alpha_s/(8\pi v)$
by using the relation $m_H^2 = 4 m_{\rm dyn}^2$ 
in the bubble approximation~\footnote{In the case of 
$m_H=m_{\rm dyn}$, we obtain 
${\cal A}_{\rm top}=(4-\pi^2/3)\cdot \alpha_s/(8\pi v) \simeq 
0.71 \cdot \alpha_s/(8\pi v)$.}. 
In any case, we obtain the ratio of ${\cal A}_{\rm col}$ and 
${\cal A}_{\rm top}$, 
\begin{equation}
 \left|\frac{{\cal A}_{\rm col}}{{\cal A}_{\rm top}}\right|
 \propto \frac{m_{\rm dyn}^2}{M^2}, 
\end{equation}
where we used $g_{t0} \approx g' \sim g_{\rm crit}$ and $g_{GAA}=g_s$.
This relation suggests that ${\cal A}_{\rm col}$ becomes comparable to 
${\cal A}_{\rm top}$ in the situation of $M \sim {\cal O}(\rm 1 TeV)$ 
and $m_{\rm dyn} \sim {\cal O}(0.6 \rm TeV)$. 

The coloron mass $M \sim {\cal O}({\rm 1 TeV})$ is favorable
in order to avoid the fine-tuning problem, 
while the large dynamical mass 
$m_{\rm dyn} \sim {\cal O}(0.6 \rm TeV)$ is obtained from a single VEV
triggering the EWSB.
The large dynamical mass can be realized within TSS models.
However, the dynamical mass is not necessarily large in TCMs with 
non-minimal Higgs sectors such as TC2.
In the next section, we estimate quantitatively
the effective $ggH$-coupling in TSS models and TC2.

\section{Size of the effective $\mbox{\boldmath $ggH$}$-coupling 
in TSS models and TC2}

\subsection{Analysis for TSS models}

In TSS models, it is possible that the EWSB occurs via 
the ``top-condensate'' only.
The large dynamical mass 
can be tuned consistently with the experimental value 
$m_t^{({\rm exp})}(=174\;\mbox{GeV})$
through the seesaw mechanism, which requires
an additional vector-like heavy quark $\chi$.
We hence consider the original version of TSS models
(the $U(1)$ tilting model)~\cite{topseesaw}, 
where the SM is embedded into a topcolor scheme,
$SU(3)_1 \times SU(3)_2 \times SU(2)_W \times U(1)_1 \times
 U(1)_2 \times U(1)_{B-L}$ gauge group. 
The gauge group $SU(3)_1 \times U(1)_1$ 
[$SU(3)_2 \times U(1)_2$] acts only 
on the third (first and second) generation(s) of quarks and 
leptons. The $U(1)_{B-L}$ charges are $x$ ($-1/3 < x < 0$)
for $t_R$ and $\chi_L$, $1/3$ for the other quarks, and $-1$ for 
leptons~\footnote{We also incorporate right-handed neutrinos.}.
The charge assignment is also shown in Table~\ref{TSS_org}.
The gauge symmetry breaking 
$SU(3)_1 \times SU(3)_2 \to SU(3)_{\rm QCD}$,
$U(1)_1 \times U(1)_2 \to U(1)_Y$, and $U(1)_{B-L}$
leave colorons, and two massive gauge bosons $Z_1$ and $Z_{B-L}$, 
respectively.
We note that the mass term $\bar{\chi}_L \chi_R$ is related to 
the $U(1)_{B-L}$ breaking VEV in the $U(1)$ tilting model.
We may take common masses $\Lambda$ for $Z_1$ and $Z_{B-L}$ 
of the order of a multi-TeV and 
the coloron mass $M$ of the order of 1 TeV.
In this situation, we can expect that 
the contribution of colorons becomes sizable and that
the constraint on the $T$-parameter can be satisfied.
Here, we comment on the Higgs boson mass $m_H$ 
in the $U(1)$ tilting model: 
it is possible that the Higgs boson has a mass of order
100 GeV with tuning of some parameters~\cite{topseesaw}.
However, the Higgs boson mass is likely
of the order of $m_{\rm dyn}$, i.e.
$m_H \sim {\cal O}(\mbox{1 TeV})$. 
We should note that the $WW$-fusion process becomes 
comparable to the gluon fusion in the mass range 
$m_H \sim {\cal O}(\mbox{1 TeV})$ in the SM.
%
%
\begin{table}[tbp]
\centering
\begin{tabular}{|c||c|c|c|c|c|c|}\hline
& $SU(3)_1$ & $SU(3)_2$ & $SU(2)_W$ & $U(1)_1$ & $U(1)_2$ 
& $U(1)_{B-L}$ \\ \hline \hline
$q_L$ & {\bf 3} & {\bf 1} & {\bf 2} & $1/6$  & $0$ & $1/3$ \\ \hline
$t_R^c$ & $\bar{\mbox{{\bf 3}}}$ & {\bf 1} & {\bf 1} & $-2/3$  
        & $0$ & $0 < -x < 1/3$ \\ \hline
$b_R^c$ & $\bar{\mbox{{\bf 3}}}$ & {\bf 1} & {\bf 1} & $1/3$  
        & $0$ & $-1/3$ \\ \hline\hline
$l_L$ & {\bf 1} & {\bf 1} & {\bf 2} & $-1/2$   & $0$ & $-1$ \\ \hline
$\tau_R^c$ & {\bf 1} & {\bf 1} & {\bf 1} & $1$ & $0$ & $1$\\ \hline
$\nu_{\tau R}^c$ & {\bf 1} & {\bf 1} & {\bf 1} & $0$ & $0$ 
                  & $1$\\ \hline\hline
$\chi_L$ & {\bf 3} & {\bf 1} & {\bf 1} & $2/3$ & $0$ 
         & $-1/3 < x < 0$ \\ \hline
$\chi_R^c$ & $\bar{\mbox{{\bf 3}}}$ & {\bf 1} & {\bf 1} 
         & $-2/3$ & $0$ & $-1/3$ \\ \hline
\end{tabular}
\caption{The representation of the third-generation and 
         $\chi$ fermion in the $U(1)$ tilting model. \label{TSS_org}}
\end{table}

Under certain conditions~\cite{topseesaw}, 
only the VEV $\VEV{\bar{t}_L \chi_R}$ acquires dynamically 
a non-zero value.
We thus refer the mass term of $\bar{t}_L \chi_R$ ($m_{t \chi}$) 
as the dynamical mass. 
The mass matrix for $t$ and $\chi$ is diagonalized by
rotating the left- and right-handed fields as follows:
\begin{subequations}
\begin{eqnarray}
 \left(\begin{array}{@{}c@{}}t_L \\ \chi_L \end{array}\right)&=&
 \left(\begin{array}{cc}c_L & s_L \\ -s_L & c_L \end{array}\right)
 \left(\begin{array}{@{}c@{}}t'_L \\ \chi'_L \end{array}\right), 
 \\
 \left(\begin{array}{@{}c@{}}t_R \\ \chi_R \end{array}\right)&=&
 \left(\begin{array}{cc}-c_R & s_R \\ s_R & c_R \end{array}\right)
 \left(\begin{array}{@{}c@{}}t'_R \\ \chi'_R \end{array}\right) ,
\end{eqnarray}
\end{subequations}
where $s_{L(R)} \equiv \sin \theta_{L(R)}, 
       c_{L(R)} \equiv \cos \theta_{L(R)}$ and 
the prime $X'$ ($X=t,\chi$) denotes the mass eigenstate.
The relations between the mass eigenvalues of $m_t$ and $m_\chi$, and 
the dynamical mass $m_{\rm dyn}$ are obtained as 
\begin{equation}
  \frac{m_t}{m_{\rm dyn}} = \frac{s_R}{c_L}, \quad
  \frac{m_{\rm dyn}}{m_\chi} = \frac{s_L}{c_R} . \label{mixing}
\end{equation}
We can determine the mixing angles $s_L,s_R$ and the dynamical mass 
$m_{\rm dyn}$ through Eq.~(\ref{mixing}) and the PS formula 
for the VEV $\VEV{\bar{t}_L \chi_R}$.
Since we obtain Yukawa vertices of $t$ and $\chi$
in mass eigenstates from
\begin{equation}
  \frac{g_{\rm t\chi}}{\sqrt{2}}
  \left( c_L \bar{t'}_L + s_L \bar{\chi'}_L\right)
  \left( s_R t'_R + c_R \chi'_R \right) H \label{yukawa}
\end{equation}
with
\begin{equation}
m_{\rm dyn}=\frac{g_{\rm t\chi}}{\sqrt{2}} v ,\quad 
g_{\rm t\chi} \equiv g_{t\chi 0} Z_H^{-1/2} , \label{m_dyn}
\end{equation}
where $g_{t \chi 0}$ is nearly equal to the critical coupling 
$g_{\rm crit}$,
we find the wave function renormalization $Z_H(p^2)$ 
of the Higgs boson with the momentum $p$ in 1-loop approximation,
%
%
\renewcommand{\arraystretch}{2.5}
\begin{table*}[tb]
 \begin{center}
\resizebox{0.95\textwidth}{!}{
\begin{tabular}[t]{|c||c|c|c|c|c|c|} \hline
$m_H$ [TeV] & 0.5 & 0.6 & 0.7 & 0.8 & 0.9 & 1 \\ \hline\hline
\quad ${\cal A}_{\rm top}$ \hfill ${\displaystyle 
\left[\frac{\times 10^{-2}}{\mbox{\footnotesize TeV}}\right]}$ 
& $1.7+1.4i$ & $1.1+1.5i$ & $0.7+1.5i$ & $0.4+1.4i$ & $0.3+1.2i$ 
& $0.1+1.1i$ \\[2mm] \hline
\quad ${\cal A}_{\rm col}$ \hfill ${\displaystyle 
\left[\frac{\times 10^{-2}}{\mbox{\footnotesize TeV}}\right]}$ 
& $-(1-3)+(0-0.1)i$ & $-(1-3)+(0.1-0.2)i$ & $-(1-3)+(0.1-0.2)i$ 
& $-(1-3)+(0.1-0.2)i$ & $-(1-3)+(0.1-0.2)i$ 
& $-(1-3)+(0.1-0.2)i$ \\[2mm] \hline
\quad ${\cal A}_{\rm heavy}$ \hfill ${\displaystyle 
\left[\frac{\times 10^{-2}}{\mbox{\footnotesize TeV}}\right]}$ 
& $\sim 0.01$ & $\sim 0.01$ & $\sim 0.01$ & $\sim 0.01$ & $\sim 0.01$ 
& $\sim 0.01$ \\[2mm] \hline \hline
\quad ${\cal A}$ \hfill ${\displaystyle 
\left[\frac{\times 10^{-2}}{\mbox{\footnotesize TeV}}\right]}$ 
& $(+1 \sim -1)+(1-2)i$ & $-(0-2)+2i$ & $-(0-2)+2i$ 
& $-(1-3)+2i$ & $-(1-3)+1i$ & $-(1-3)+1i$ \\[2mm] \hline
  \end{tabular}
}  \caption{Sizes of ${\cal A}_{\rm top}$, ${\cal A}_{\rm col}$ and 
            ${\cal A}_{\rm heavy}$ for various $m_H$, 
            obtained from Eqs.~(\ref{A_top})--(\ref{A_col}). 
            The $H \to gg$ amplitude ${\cal A}$ is given by
            ${\cal A}={\cal A}_{\rm top} + {\cal A}_{\rm heavy} 
                                         + {\cal A}_{\rm col}$.
            We have used the values for $s_L,s_R$ and $m_{\rm dyn}$ 
            as shown in Eq.~(\ref{dyn}) with 
            $\alpha_s=0.12$, $M=2.1$ TeV, 
            $\Lambda=20-30$ TeV and $m_\chi=5-10$ TeV.
            \label{tab_A}}
 \end{center}
\end{table*}
\renewcommand{\arraystretch}{1}
\begin{widetext}
\begin{eqnarray}
Z_H (p^2)&=&
 -\frac{N_c g_{t\chi 0}^2}{8\pi^2}\left( \frac{1}{2}+
  3(s_L^2 s_R^2+c_L^2 c_R^2)\int_0^1 dx x(1-x)\log\left[
    \frac{D_2(m_\chi^2,m_t^2;p^2)}{\Lambda^2+D_2(m_\chi^2,m_t^2;p^2)}
    \right] \right. \nonumber \\
&& \qquad
 +3s_L^2 c_R^2 \int_0^1 dx x(1-x)\log\left[
    \frac{D_2(m_\chi^2,m_\chi^2;p^2)}
         {\Lambda^2+D_2(m_\chi^2,m_\chi^2;p^2)} \right] \nonumber \\
&& \qquad \left.
 +3c_L^2 s_R^2 \int_0^1 dx x(1-x)\log\left[
    \frac{D_2(m_t^2,m_t^2;p^2)}{\Lambda^2+D_2(m_t^2,m_t^2;p^2)}
    \right]
  \right) ,
\end{eqnarray}
where we defined
\begin{equation}
  D_2(M_1^2,M_2^2;p^2) \equiv xM_1^2+(1-x)M_2^2-x(1-x)p^2,
\end{equation}
and we used the regularization with the naive ultraviolet-cutoff 
$\Lambda$. 
Noting that $\Lambda^2, m_\chi^2 \gg m_t^2, p^2$, 
we can obtain a more convenient expression for $Z_H$,
\begin{eqnarray}
  Z_H(p^2) &\simeq& \left.\frac{N_c g_{t\chi 0}^2}{(4\pi)^2}\right[\,
  (s_L^2+c_L^2 c_R^2)\log (1+\Lambda^2/m_\chi^2) \nonumber \\
&& \quad \left.
 +c_L^2 s_R^2\left\{\log\Lambda^2/m_t^2+\frac{5}{3}+\frac{4m_t^2}{p^2}
   -\left(1+\frac{2m_t^2}{p^2}\right)F(m_t^2,p^2)\right\}+k\,\right]
\label{Z_H}
\end{eqnarray}
with 
\begin{equation}
F(m^2,p^2)\equiv \left\{\begin{array}{lc}
 \sqrt{1-\frac{4m^2}{p^2}}\log\frac{1+\sqrt{1-\frac{4m^2}{p^2}}}
                                  {1-\sqrt{1-\frac{4m^2}{p^2}}}
 -i\pi\sqrt{1-\frac{4m^2}{p^2}} & (p^2 > 4m^2),\\
 2\sqrt{\frac{4m^2}{p^2}-1}\arctan\frac{1}{\sqrt{\frac{4m^2}{p^2}-1}}
& (p^2 \leq 4m^2), \end{array} \right. 
\end{equation}
where we introduced a constant term $k(\approx 1)$ 
arising from ambiguities of various regularizations.
Substituting Eq.~(\ref{Z_H}) with $p^2=0$ for Eq.~(\ref{m_dyn}),
we obtain the PS formula,
\begin{equation}
\frac{v^2}{m_t^2}\frac{s_R^2}{c_L^2} = \frac{N_c}{8\pi^2}\left[\,
(s_L^2 + c_L^2 c_R^2)\log (1+\Lambda^2/m_\chi^2) + 
c_L^2 s_R^2 \log \Lambda^2/m_t^2+k\,\right]. \label{PS}
\end{equation}
\end{widetext}
Since we find that Eq.~(\ref{PS}) is a closed equation 
for $s_R$ due to the relations (\ref{mixing}),
$(s_L s_R)/(c_L c_R) = m_t/m_\chi$, 
we can solve Eq.~(\ref{PS}) numerically, 
\begin{equation}
  m_{\rm dyn}=0.5-0.9\mbox{ TeV}, \quad s_R=0.2-0.3, \quad 
  s_L \simeq 0.1 \label{dyn}
\end{equation}
with $v=246$ GeV and $m_t=174$ GeV, where we used
$k=0,1,2$, $\Lambda=20-30$ TeV and $m_\chi=5-10$ TeV,
for example.
These parameters are fairly safe from 
the present constraint on the $T$-parameter 
with $m_H=$1 TeV~\cite{Marciano:2000yj}. 

Now, we estimate the $H \to gg$ amplitude ${\cal A}$.
Using Eq.~(\ref{yukawa}),  the contribution of the top-quark 
loop is given by
\begin{equation}
{\cal A}_{\rm top}({\rm TSS}) = c_L^2 {\cal A}_{\rm SM}(\tau) ,
\label{A_top}
\end{equation}
which is almost equivalent to the SM contribution.
Since we consider the Higgs boson mass 
to be larger than the threshold of the top-pair production, 
${\cal A}_{\rm top}(\rm TSS)$ has an imaginary part.
(See also Eqs.~(\ref{sm_finite}) and (\ref{f}).)
Eqs.~(\ref{mixing}), (\ref{yukawa}) and (\ref{m_dyn}) lead to 
the contribution of the $\chi$ field
\begin{equation}
  {\cal A}_{\rm heavy}({\rm TSS})
 =s_L c_R \frac{m_{\rm dyn}}{m_\chi}\frac{\alpha_s}{12\pi v}
 =s_L^2 \frac{\alpha_s}{12\pi v}
  \label{A_chi}
\end{equation}
at the leading order of $m_H^2/m_\chi^2$.
Although we do not explicitly integrate out the heavy fermion $\chi$, 
this effect is translated in the estimate of 
${\cal A}_{\rm heavy}(\rm TSS)$ in Eq.~(\ref{A_chi}).
We evaluate the coloron contribution ${\cal A}_{\rm col}$ 
in TSS models from Eq.~(\ref{ggH}),
\begin{eqnarray}
  {\cal A}_{\rm col}({\rm TSS}) &=& 
 -\frac{N_c^{-1} (g_{GAA}+g_s)^2}{4Z_H(m_H^2)}
  \frac{g'{}^2}{g_{t\chi 0}^2}\frac{v}{M^2} \nonumber \\ &\sim&
 -\frac{\alpha_s}{\pi v}\frac{m_{\rm dyn}^2}{M^2}
  \frac{Z_H(0)}{Z_H(m_H^2)}  \label{A_col}
\end{eqnarray}
with $p^2=m_H^2$, $g_{GAA}=g_s$, 
$g_{t \chi 0} \approx g' \sim g_{\rm crit}$ and 
$N_c g_{t \chi 0}^2 \sim 8\pi^2$.
We note that ${\cal A}_{\rm col}(\rm TSS)$ has an imaginary part 
arising from $Z_H(m_H^2)$ due to $m_H^2 > 4 m_t^2$.
Numerically, we find
${\cal A}_{\rm top},{\cal A}_{\rm col}$ and ${\cal A}_{\rm heavy}$,
\begin{eqnarray}
{\cal A}_{\rm top}({\rm TSS})&=&
   \frac{\left[(1.7-0.1)+(1.5-1.1)i\right]\times 10^{-2}}
        {\mbox{(1 TeV)}}, \\
{\cal A}_{\rm col}({\rm TSS})&=&
   \frac{\left[-(1-3)+i(0-0.2)\right]\times 10^{-2}}
        {(1\;{\rm TeV})}
\end{eqnarray}
and
\begin{equation}
  {\cal A}_{\rm heavy}({\rm TSS}) \sim
   \frac{10^{-4}}{(1\;{\rm TeV})} ,
\end{equation}
for $m_H=0.5-1$ TeV with $\alpha_s=0.12$, 
$\Lambda=20-30$ TeV and $m_\chi=5-10$ TeV, where we used 
the reference value of the coloron mass $M=2.1$ TeV, 
corresponding to the expected bound for the direct production 
at TeV~33~\cite{harris}.
(See also Table~\ref{tab_A}.)
The $H \to gg$ amplitude ${\cal A}$ is given by 
${\cal A}={\cal A}_{\rm top}+{\cal A}_{\rm heavy}+{\cal A}_{\rm col}$
and found numerically,
\begin{equation}
  {\cal A}({\rm TSS})=
  \frac{\left[(+1 \sim -3)+(1-2)i\right] \times 10^{-2}}
       {(1\;{\rm TeV})} ,
\end{equation}
for $m_H=0.5-1$ TeV.
Although the real part of ${\cal A}_{\rm top}$ is comparable 
to the size of its imaginary part up to $m_H=0.6$ TeV, 
the imaginary part becomes dominant around $m_H=1$ TeV. 
The real part of ${\cal A}_{\rm col}$ is sizable for $m_H=0.5-1$ TeV,
while the imaginary part of ${\cal A}_{\rm col}$ is negligible.
The contribution ${\cal A}_{\rm heavy}$ of the $\chi$ field 
is also suppressed due to the constraint on the $T$-parameter 
($s_L \ll 1$).
We thus find that the real (imaginary) part of 
the $H \to gg$ amplitude ${\cal A}$ is dominated by 
${\cal A}_{\rm col}$ (${\cal A}_{\rm top}$) 
for $m_H=0.8-1$ TeV. 

We have shown that the contribution of colorons can be sizable in 
the $U(1)$ tilting model in the situation of 
$M \sim {\cal O}({\rm 1 TeV})$ and 
$\Lambda \sim m_\chi \sim {\cal O}({\rm 10 TeV})$.
However, we have not included higher order loop corrections.
Since colorons are strongly coupled to the top quark,
diagrams including top-quark loops 
are possibly sizable.
In TSS models, the amplitude of $H \to V_L V_L (V_L=W_L, Z_L)$ is 
also likely to receive non-perturbative effects due to 
the heavy Higgs boson mass near the perturbative unitarity bound, 
$m_H \sim {\cal O}(\mbox{1 TeV})$. 
Thus, the Higgs boson in TSS models may be studied 
in non-perturbative approaches such as Bethe-Salpeter equations.
This, however, is out of the scope of this paper. 
An analysis of the Higgs boson production of TSS models
at the LHC will be performed elsewhere~\cite{future}.

\subsection{Analysis for TC2}

%
%
\renewcommand{\arraystretch}{2.0}
\begin{table*}[tbp]
  \begin{center}
   \begin{tabular}{|l|c|c|} \hline
   \hfil $m_H$ [GeV] \hfil & 180 & 190 \\ \hline\hline
   signals of $H$ in TC2 [fb] & $8-14$ & $7-12$ \\ \hline
   signals in the SM [fb] & $0.85$ & $0.73$ \\ \hline
   backgrounds [fb] & $3.8$ & $7.5$ \\ \hline
   $S/\sqrt{B}$ for 2 fb$^{-1}$ in TC2 & $6-10$ & $3-6$ \\ \hline
   $S/\sqrt{B}$ for 2 fb$^{-1}$ in the SM & $0.62$ & $0.38$ \\ \hline
   \end{tabular}
  \caption{Expected signals of
   $gg \to H \to W^* W^* \to \ell \bar{\ell} \nu \bar{\nu}$
   in TC2 after the kinematical cuts and the likelihood cut 
   at the Tevatron.
   In the table, we estimate signals of the Higgs boson in TC2
   by using Eq.~(\ref{A_TC2})
   and the SM value shown in Table 29 in Ref.~\cite{tevatron}.
   There is no study of backgrounds for $m_H \geq 200$ GeV 
   in Ref.~\cite{tevatron}.
   We assume that the main decay mode of the Higgs boson in TC2 is 
   pair production of weak bosons. \label{tab1}}
  \end{center}
\end{table*}

\begin{table*}[tbp]
  \begin{center}
   \begin{tabular}{|l|c|c|c|c|} \hline
   \hfil $m_H$ [GeV] \hfil & 200 & 240 & 280 & 320\\ \hline\hline
   signals of $H$ in TC2 for 30fb$^{-1}$ & 
   $486-864$ & $792-1408$ & $810-1440$ & $810-1440$ \\ \hline
   signals in the SM for 30fb$^{-1}$ & 
   $54$ & $88$ & $90$ & $90$ \\ \hline
   backgrounds for 30fb$^{-1}$ & $7$ & $15$ & $17$ & $16$ \\ \hline
   $S/\sqrt{B}$ for 30fb$^{-1}$ in TC2 
   & $184-327$ & $204-364$ & $196-349$ & $203-360$\\ \hline
   $S/\sqrt{B}$ for 30fb$^{-1}$ in the SM
   & $20.4$ & $22.7$ & $21.8$ & $22.5$\\ \hline
   \end{tabular}
  \caption{Expected signals of $H \to ZZ \to 4\ell$
   in TC2 after $p_T$ cut at the LHC.
   In the table, we estimate signals of the Higgs boson in TC2
   by using Eq.~(\ref{A_TC2})
   and the SM value shown in Table 19-21 in Ref.~\cite{LHC}.
   We assume that the main decay mode of the Higgs boson in TC2 is 
   pair production of weak bosons. \label{tab2}}
  \end{center}
\end{table*}

In TC2, the condensate of the techni-fermion $T$ mainly triggers
the EWSB,
\begin{equation}
  v^2=v_{\rm TC}^2 + v_{\rm top}^2, \quad v_{\rm TC} \sim v,
\end{equation}
where condensates of $\bar{T}T$ and $\bar{t}t$ provide
$v_{\rm TC}$ and $v_{\rm top}$, respectively.
Adjusting the value of $v_{\rm top}$, we can obtain 
the experimental value of the top quark mass $m_t^{({\rm exp})}$:
the PS formula,
\begin{equation}
  v_{\rm top}^2 = \frac{N_c}{8 \pi^2}
  (m_t^{({\rm exp})})^2 \ln M^2/(m_t^{({\rm exp})})^2,
\end{equation}
which is same as Eq.~(\ref{PS0}),
leads to $v_{\rm top}/v=1/(3-4)$. 
While the dynamical mass is small in TC2,
$m_{\rm dyn}=m_t^{({\rm exp})}=174$ GeV, 
the top-loop effect becomes very large 
due to enhancement of the top-Yukawa coupling 
($g_t \propto m_t^{({\rm exp})}/v_{\rm top}$).
As a result, the gluon fusion process does not suffer from 
the contribution of colorons.
We estimate the Higgs boson mass in TC2 as 
$m_t^{({\rm exp})} < m_H < 2m_t^{({\rm exp})}$.

We easily obtain the contribution of colorons 
${\cal A}_{\rm col}$ in TC2 from Eq.~(\ref{ggH2}): 
\begin{equation}
{\cal A}_{\rm col}({\rm TC2}) \sim
 -\frac{\alpha}{\pi v_{\rm top}}
         \frac{(m_t^{({\rm exp})})^2}{M^2} 
 \sim - \frac{0.3 \times 10^{-2}}{(\rm 1\;TeV)}
\end{equation}
with $g_{t0} \approx g' \sim g'_{\rm crit}$ and $M=2.1$ TeV.
On the other hand, the effect of enhancement of the top-Yukawa 
coupling is given by
\begin{equation}
{\cal A}_{\rm top}({\rm TC2}) = \frac{v}{v_{\rm top}}
{\cal A}_{\rm SM}(\tau), \quad \frac{v}{v_{\rm top}} \simeq 3-4, 
\end{equation}
where the value of the SM is numerically found,
\begin{equation}
 {\cal A}_{\rm SM}(\tau) = 
 \frac{(1.4-1.7) \times 10^{-2}}{(\rm 1\;TeV)},
\end{equation}
for $m_H=180-320$ GeV.
We emphasize that the value of $v/v_{\rm top}$ is not changed, 
even if the coloron mass $M$ is taken to the order of a multi-TeV.
Unless we specify the matter content of TC2 in detail,
we cannot estimate contributions of techni-fermions having 
QCD charges, i.e. ${\cal A}_{\rm heavy}$ in our notation. 
For example, techni-fermions in models of Ref.~\cite{TC2}
do not have QCD charges, 
so that they do not contribute to the gluon fusion process. 
We then estimate the enhancement factor to the SM value as
\begin{equation}
 {\cal A}({\rm TC2}) = {\cal A}_{\rm top}+{\cal A}_{\rm col}
 \sim (3-4) \times {\cal A}_{\rm SM} \label{A_TC2}
\end{equation}
with ${\cal A}_{\rm heavy}=0$
and find that the coloron effect ${\cal A}_{\rm col}$ is negligible 
compared to the contribution of the top-loop ${\cal A}_{\rm top}$.

Now, we study the signature of Higgs boson production 
at hadron colliders.
Since masses of top-pions $\pi_t$ much below $\sim 165$ GeV are 
phenomenologically forbidden~\cite{t2bt_pi} due to the absence of 
the decay mode $t \to \pi_t^+ + b$, 
the main decay mode of the Higgs boson is expected 
to be pair production of weak bosons.
In this situation, we can apply the SM analysis 
at the Tevatron~\cite{tevatron} and at the LHC~\cite{LHC} directly 
to TC2.
Although $HWW$- and $HZZ$-couplings in TC2 are suppressed,
\begin{equation}
  g_{HVV}({\rm TC2})/ g_{HVV}({\rm SM})=v_{\rm top}/v, \quad V=W,Z
\end{equation}
the effects are irrelevant 
under the narrow width approximation, because
the branching ratio of $H \to VV$ is not changed from 
the SM in the mass range $m_t^{({\rm exp})}< m_H <2m_t^{({\rm exp})}$.
The considerable enhancement of the top-Yukawa gives a chance 
to observe the Higgs boson in TC2
even at the Tevatron Run II. (See Table~\ref{tab1}.)
While we cannot expect to find any evidence of 
the SM Higgs boson with $m_H \sim 200$ GeV 
for an integrated luminosity of 2~fb$^{-1}$ at the Tevatron Run II,
we can estimate $S/\sqrt{B}=3-6$ for the Higgs boson in TC2
with 2~fb$^{-1}$ and $m_H=190$ GeV.
We show expected signals of the Higgs boson in TC2 
up to $m_H=190$ GeV at the Tevatron, 
since there is no background estimate for $m_H \geq 200$ GeV 
in Ref.~\cite{tevatron}.
At the LHC, signals of the Higgs boson in TC2 are considerably 
enhanced. (See Table~\ref{tab2}.)
The Higgs boson in TC2 with the mass range 
$m_t^{({\rm exp})}< m_H <2m_t^{({\rm exp})}$ 
can be discovered at the LHC much more easily than the SM Higgs boson.

\section{Summary and discussion}

In this paper, we have studied the effective $ggH$-coupling in TCMs.
We have considered the effect of colorons, in addition to
the loop contributions of the top quark and other heavy quarks.
In order to estimate the coloron effect, we have derived 
the low-energy effective theory by eliminating colorons 
by using the EOM for colorons.
We have found that the contribution of colorons ${\cal A}_{\rm col}$ 
is proportional to $m_{\rm dyn}^2/M^2$.
Thus, ${\cal A}_{\rm col}$ becomes sizable for 
the coloron mass $M \sim {\cal O}(\rm 1 TeV)$ and 
the dynamical mass $m_{\rm dyn} \sim {\cal O}(0.6 \rm TeV)$. 
An important point is that the large dynamical mass 
$m_{\rm dyn} \sim {\cal O}(0.6 \rm TeV)$ can be realized 
consistently with the experimental value of the top-quark mass 
$m_t^{(\rm exp)}$ in TSS models, 
while the dynamical mass itself is adjusted to $m_t^{(\rm exp)}$
in TC2.
We have shown that the contribution of colorons is actually sizable 
in TSS models with $M \sim {\cal O}(\rm TeV)$ and 
$m_\chi \sim {\cal O}(10 \rm TeV)$: 
we have evaluated the contributions of colorons and the top-loop as
${\cal A}_{\rm col}=[-(1-3)+(0.1-0.2)i]\times 10^{-2}
(\rm TeV^{-1})$ and 
${\cal A}_{\rm top}=[(0.4-0.1)+(1.4-1.1)i]\times 10^{-2}
(\rm TeV^{-1})$, which is almost same as the SM value,
for $m_H=0.8-1$~TeV with some parameters 
being safe from the constraint on the $T$-parameter.
Namely, ${\cal A}_{\rm col}$ (${\cal A}_{\rm top}$) is
numerically dominant in the real (imaginary) part of 
the $H \to gg$ amplitude for $m_H \sim {\cal O}(\rm 1 TeV)$. 
In TC2, we have found that the effect of colorons is negligible 
compared to the contribution of the top-loop,
because the dynamical mass $m_{\rm dyn}$ itself is small and 
the top-Yukawa coupling is enhanced by a factor of $3-4$.
Since the Higgs boson production is considerably enhanced in TC2,
we can obtain signatures of the Higgs boson in TC2 with 
$m_H \sim 200$ GeV even at the Tevatron Run II as well as at the LHC.
In particular, we have evaluated $S/\sqrt{B}=3-6$ 
for an integrated luminosity of 2 fb$^{-1}$ 
and $m_H=190$ GeV at the Tevatron Run II.

We have not studied the expected signals of the Higgs boson 
in TSS models at hadron colliders. 
Since the $WW$-fusion process becomes comparable 
to the gluon fusion in the mass range 
$m_H \sim {\cal O}(\mbox{1 TeV})$, 
we would have also needed to study the $WW$-fusion process 
for TSS models. 
In addition, we may investigate the Higgs boson production in 
non-perturbative approaches, 
because $m_H \sim {\cal O}(\mbox{1 TeV})$ is near the perturbative
unitarity bound.
A detailed analysis of the expected signals of the Higgs boson 
in TSS models will be performed elsewhere~\cite{future}.

\section*{Acknowledgments}
The author is very grateful to M.Tanabashi for helpful discussions.

\end{document}